\documentclass[useAMS,usenatbib,usegraphicx]{mn2e}
\usepackage{subfig}
\include{epsf}

\title[Towards a self-consistent numerical galaxy model]
{Towards a self-consistent numerical model of late-type galaxies: Calibrating the effects of sub-grid physics on galactic models}
\author[Rahimi et~al.]
 {Awat Rahimi,$^{1}$\thanks{E-mail: ara2@mssl.ucl.ac.uk}
Daisuke Kawata,$^{1}$
\\
$^{1}$ Mullard Space Science Laboratory, University College London,
Holmbury St. Mary, Dorking, Surrey, RH5 6NT
\\
}
\date{Accepted .
      Received ;
      in original form }

\pagerange{\pageref{firstpage}--\pageref{lastpage}}
\pubyear{2011}

\begin{document}

\maketitle

\label{firstpage}

\begin{abstract}
We carry out several isolated galaxy evolution simulations in a fixed dark matter halo gravitational potential using the new version of our N-body/Smoothed Particle Hydrodynamics (SPH) code {\tt GCD+}. The new code allows us to more accurately model and follow the evolution of the gas and stellar components of the system including powerful supernovae feedback and its effects on the inter-stellar medium  (ISM). 
Here we present the results of six simulations of an M33-sized late-type disc galaxy each with varying values for our model parameters which include the star formation efficiency $(C_{\rm *})$, the energy released per supernovae explosion ($E_{\rm SN}$) and the energy released per unit time from stellar winds ($E_{\rm SW}$). We carry out both a pixel-by-pixel and radial ring analysis method for each of our galaxies comparing our results to the observed Schmidt-Kennicutt Law and vertical gas velocity dispersion versus radius relation amongst others. We find that our models with higher feedback more closely resemble the observations and that feedback plays a pivotal role in obtaining {\it both} the observed Schmidt-Kennicutt and gas velocity dispersion relations. 
\end{abstract}


\section{Introduction}
\label{intro-sec}

The process of galaxy formation and evolution remains one of the most outstanding problems in modern astrophysics. From a cosmological point of view, great progress has been made including the development of a theory; $\Lambda$CDM cosmology which accurately explains the present day large scale distribution of galaxies from the primordial small scale density fluctuations in the early universe. The actual processes which govern individual galaxy formation and evolution are far less well understood however. One of these important processes is that of star formation itself. \citet{sch59} suggested that the star formation rate (SFR) density and the interstellar gas density in our Galaxy were related by a simple power law relation of the form $\rho _{\rm SFR}\propto (\rho _{\rm gas})^{n}$, where $n$ is the power law index. \citet{sch59} originally suggested a value for $n$ of 2, so that the rate of star formation varies with the square of the gas density. Studying the form of the observed star formation law,
\citet{rk89,ken98} analysed the SFR {\it surface} density and the gas {\it surface} density  
 for a much larger sample of galaxies and over a wider range of gas densities, determining the  power law index $n$ to be equal to 1.4 \citep[see also][]{ken98AR}. We now refer to the empirical relation between SFR and gas surface density as the Schmidt-Kennicutt (SK) relation. Recently, the SK relation has been studied in much more depth \citep[e.g.][]{blw08,lwb08} including observing the relation for individual components of the gas (HI and molecular) and as a function of environment. 
Several observational studies found evidence for a two-slope SK relation \citep[e.g.][]{blw08}.
 \citet{blw08} showed that a break exists at gas surface densities of 9 M$_{\odot}\textup{yr}^{-1}$, below which the slope of the SFR steepens. This break corresponds to the transition between the HI and molecular phase of the gas. Theoretical models also predict such breaks in the SK relation. For example \citet{kmt09} made a theoretical model consisting of three prescriptions: molecular self-shielding, internal feedback and turbulent regulation in molecular clouds which address the physics responsible for determining the SFR on local scales and successfully managed to reproduce the observed trends in the SK relation.

The SK relation is of great importance as it is used for the calibration of numerical codes of galaxy formation and evolution, with the SFR often explicitly required to obey a SK type law \citep[e.g.][]{sh03,bto07,sd08}. Some more recent studies however, do not force the SFR to follow a SK-like relation, and so its adherence to this law is a prediction of the model \citep[e.g.][]{rk08,sdk08,mmb11}.

Recently \citet{tam09} showed that the gas velocity dispersion is also an important factor to consider for any galaxy evolution model. Using HI line width maps obtained  for 11 disc galaxies of The HI Nearby Galaxy Survey \citep[THINGS,][]{wbb08} they determined that all of their sample galaxies showed a systematic radial decline in their HI velocity dispersion, with a characteristic value of $\sim$ 10 km s$^{-1}$ by $r_{25}$ $-$ roughly the radial extent of the star forming disc. The radial decline in velocity dispersion was well correlated with the radial decline of SFR and therefore they suggested that high gas velocity dispersion is induced by supernovae feedback.

The implications of such studies are that the gas velocity dispersion should now be used as an additional constraint for any new numerical simulation codes of galaxy evolution. Recently, \citet{dbp11} performed isolated disc galaxy evolution simulations to investigate how the properties of the ISM, molecular clouds and SFR depend on the level of stellar feedback, comparing their results against both the observed SK relation and gas velocity dispersion. 
They found that supernovae were vital in reproducing the observed velocity dispersions and scale heights of the individual components of the ISM.
Recently, \citet{pgc11} analysed both the SK and HI velocity dispersion relations for a fully consistent cosmologically simulated late-type dwarf spiral galaxy. 

In this paper we initiate a project to create an accurate and self-consistent numerical model for late-type spiral galaxies. To this end we have been developing our numerical simulation code {\tt GCD+} \citep{kg03a}.
The new version of the code includes new features and improvements to a number of aspects of the modelling of galaxy formation and evolution, including new star formation and feedback recipes. The overall effect of these changes is that the code can now capture and accurately follow the effects of powerful supernovae explosions on the evolution of the galaxy. The details of the new model are described in the next section.
In this study, we set up an isolated disc galaxy consisting of gas and stellar discs with no bulge component, in a static dark matter halo gravitational potential. 
To calibrate our new model, we here focus on M33{\it-like} bulgeless galaxies.
M33 is a suitably chosen late-type bulgeless spiral galaxy. Its close proximity in the Local Group means it has been a favourite choice for detailed observational campaigns \citep[e.g.][]{hcs04,vcg10}. Its smaller size compared to the Milky Way also means that we can simulate it to much higher resolution using the same number of particles. M33 also lacks a supermassive black hole at its centre \citep{mfj01}, a factor which helps simplify the modelling and analysis.
Therefore, M33{\it-like} galaxies are ideal target galaxies for testing and calibrating the numerical modelling of the physical processes in galaxy evolution.
Note that our aim is $\it{not}$ to reproduce completely the observed properties of M33.
It is likely that M33 had a recent interaction with M31 \citep{mii09,ppm09} and so it may not be a typical galaxy. We focus more on the {\it general} observed properties for M33{\it-sized} disc galaxies, such as the SK relation and the gas velocity dispersion.

The paper is organised as follows. In Section 2, we describe the galaxy model together with  some new features of our numerical simulation code. In Section 3, we present our results and analyses. We summarise our findings and present our conclusions and planned future work in Section 4.

\begin{table*}
 \centering
 \begin{minipage}{140mm}
  \caption{Simulation parameters}
  \renewcommand{\footnoterule}{}
  \begin{tabular}{@{}llllllllll@{}}
  \hline
   $C_{\rm NFW}$ & $M_{\rm vir}$ & $r_{\rm vir}$ & $M_{\rm Star}$ & $M_{\rm Gas}$ 
   & $R_{\rm d,star}$ & $R_{\rm d,gas}$ & $z_{\rm d,star}$ \\
   & (M$_{\odot})$ & (kpc) & (M$_{\odot})$ & (M$_{\odot})$ & (kpc) & (kpc) & & \\
   \hline
10 & $5.0\times10^{11}$ & 163 & $4.0\times10^{9}$ & $2.0\times10^{9}$ & 1.4 & 4.0 & 0.3  \\
 \hline
\end{tabular}
\end{minipage}
\end{table*}

\section{The code and model}
\label{code-sec}

\subsection{New version of {\tt GCD+}}

To simulate our galaxy, we use an updated version of our original galactic chemodynamical evolution code {\tt GCD+} \citep{kg03a}. 
{\tt GCD+} is a parallel three-dimensional N-body/SPH \citep{ll77,gm77} code which can be used in studies of galaxy formation and evolution in both a cosmological and isolated setting.
{\tt GCD+} incorporates self-gravity, hydrodynamics, radiative cooling, star formation, supernova feedback, and metal enrichment. The new version of the code includes metal diffusion suggested by Greif et al. (2009).
Our recent update and performance in 
various test suites will be presented in a series of our forthcoming papers (Kawata et al. in preparation). 
Here we briefly describe the major points of our recent upgrade. 
We have implemented a modern scheme of SPH suggested by \citet{rp07}. We follow
their artificial viscosity switch \citep{jm97} and artificial thermal conductivity to resolve
the Kelvin-Helmholtz instability \citep{koc09}. Following \citet{sh02}, instead of the energy equation, we integrate the entropy equation. As suggested by \citet{sm08}, we have added the individual time step limiter, which is crucial for correctly resolving the expansion bubbles induced by supernovae feedback \citep[see also][]{mbg10,ddv11}. We also implement the FAST scheme \citep{sm10} which allows the use of different timesteps for integrating hydrodynamics and gravity. The code now also includes adaptive softening for both the stars and gas \citep{pm07} although in this current study we only implement it for the gas particles (see below).

 Radiative cooling and heating is calculated with CLOUDY \citep[v08.00:][]{fkv98}. We tabulate cooling and heating rates and the mean molecular weight as a function of redshift, metallicity, density and temperature adopting the 2005 version of the \citet{hm96} UV background radiation.
We also add a thermal energy floor following \citet{rk08} to keep the Jeans mass higher than $2N_{\rm nb} m_{\rm p}$, where $N_{\rm nb}$ is the number of neighbour particles and $m_{\rm p}$ is the particle mass, to avoid numerical instability \citep[see also][]{bb97}. In the resolution of simulations presented in this paper (see below), the gas particle meets this condition around $n_{\rm H}=1$ cm$^{-3}$. Therefore, we allow the gas particle to become a star particle if the density becomes greater than $n_{\rm H}=1$ cm$^{-3}$, i.e. the star formation threshold density, and if their velocity field is convergent, following the Schmidt law as described in \citet{kg03a}:
\begin{equation}
 \frac{d \rho_*}{dt} = -\frac{d \rho_{\rm g}}{dt}
 = \frac{C_* \rho_{\rm g}}{t_{\rm g}},
\label{sfreq}
\end{equation}
where $C_*$ is a dimensionless SFR parameter and $t_{\rm g}$ is the dynamical time.

 We assume that stars are distributed according to the \citet{s55} initial mass function (IMF). {\tt GCD+} takes into account chemical enrichment by both Type II supernovae \citep[SNe II,][]{ww95} and Type Ia supernovae \citep[SNe Ia,][]{ibn99,ktn00} and mass loss from intermediate-mass stars \citep{vdhg97}, and follows the chemical enrichment history of both the stellar and gas components of the system. The new version of {\tt GCD+} uses a different scheme for star formation and feedback. We now keep the mass of the baryon (gas and star) particles completely the same, unlike \citet{kg03b} or the majority of  SPH simulations which include star formation. Although the basic strategy is similar to \citet{lpc02} and \citet{mss02}, a slightly different implementation is adopted. Details of the methodology will be described in our forthcoming paper. Here we describe it briefly. First, every star particle formed in the simulation is randomly assigned a mass group ID ranging from 1 to 61 (although 61 is chosen arbitrarily, it is a compromised selection to sample the stellar mass range and be similar to our resolution, i.e. number of neighbour particles). We can calculate that about 13 per cent of the mass is ejected by SNe II from a star cluster following the assumed IMF and \citet{ww95}. Therefore we set ID=1-8 particles to be `SNe II particles'. Each SNe II particle receives 1/8 of the energy and metals produced by SNe II depending on the age and initial metallicity. After the age reaches a time when 8/61 of the mass is ejected (about the lifetime of a 8 M$_{\odot}$ star, depending on metallicity), the whole star particle is changed back to a gas particle. We assume each supernova produces thermal energy $E_{\rm SN}$ (ergs). We also assume that stellar winds from massive stars ($>30$ M$_{\odot}$) produce thermal energy $E_{\rm SW}$ (ergs s$^{-1}$) and add this to the SNe particles. We explore the effect of these parameters applying $E_{\rm SN}=10^{50}$ and $10^{51}$ ergs and $E_{\rm SW}=10^{36}$ and $10^{37}$ ergs~s$^{-1}$ \citep[e.g.][]{bg94}. Consequently SNe II particles have higher temperature and become metal rich. We do not take into account radiative cooling until the particle turns back to a gas particle. However, we calculate thermal energy for the SNe II particles following the SPH scheme and add the pressure from the SNe II particles to their neighbour particles, while the dynamics of the SNe II particles are only affected by gravity \citep[see also][]{pvi04}. The metals produced by SNe II are also distributed from the SNe II particles through the metal diffusion scheme of \citet{ggb09}, i.e. the metal diffusion applied to the SNe II particles is the same as the other normal gas particles. We apply a similar algorithm to star particles with different IDs. From the IMF about 30 per cent of the mass will be ejected after a cosmic time. Therefore star particles with IDs ranging from 9 to 19 will turn back into a gas particle depending on the age of the star particle, which models the mass loss from intermediate mass stars and SNe Ia feedback. For these particles, we turn on radiative cooling during their mass loss and SNe Ia. For example,  ID 9 particles with solar metallicity represent mass-loss from 7 M$_{\odot}$ to 5.6 M$_{\odot}$ stars. When their age becomes older than the lifetime of a 7 M$_{\odot}$ star, they become a `feedback particle'. The particle inherits the original metal abundance and receives additional metals that stars in this mass range produce \citep{vdhg97}, which are then diffused via the metal diffusion scheme applied. Their thermal energy is calculated with the SPH scheme, and the additional pressure from these feedback particles are applied to their neighbouring particles (especially for the higher ID particles with SNe Ia feedback). Once the particle becomes older than the lifetime of a 5.6 M$_{\odot}$ star, the particle becomes a normal gas particle. Due to this algorithm, the particle mass of stars and gas are always constant, and the mass resolution is kept constant. 
 
  The main free parameters of the new version of {\tt GCD+} include the star formation efficiency, $C_*$, energy per supernova, $E_{\rm SN}$, and stellar wind energy per massive star, $E_{\rm SW}$. Especially these parameters control the effect of feedback on star formation, which is the most unknown and possibly most important process in galaxy formation and evolution. This paper explores the effect of these parameters and calibrates these parameters against the observed properties, such as the SK relation and the gas velocity dispersion.
  
\begin{figure*}

\centering
   
\subfloat{\label{fig:Run1}\includegraphics[scale=0.34]{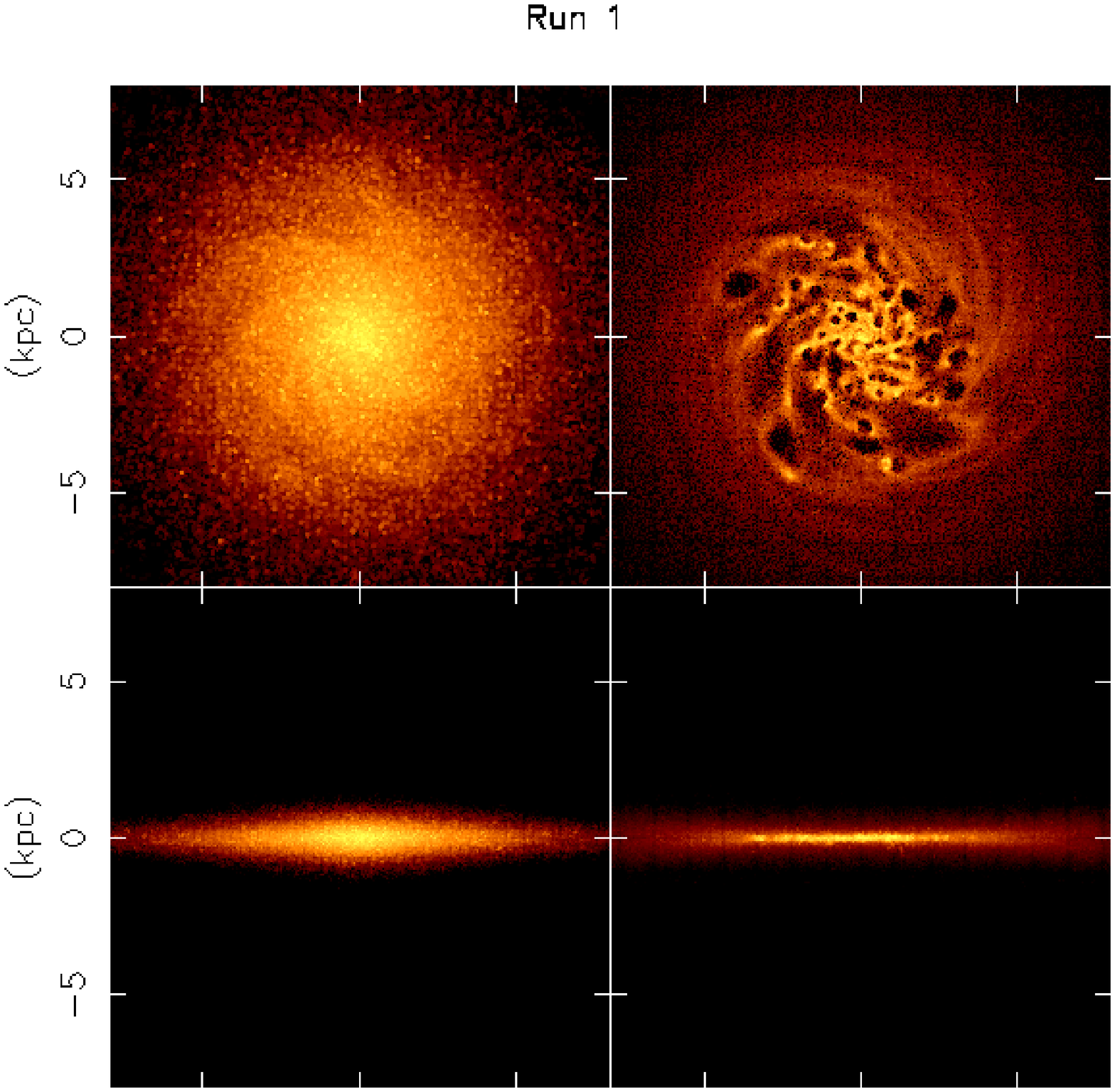}} \hspace{5.0mm}   
\subfloat{\label{fig:Run2}\includegraphics[scale=0.34]{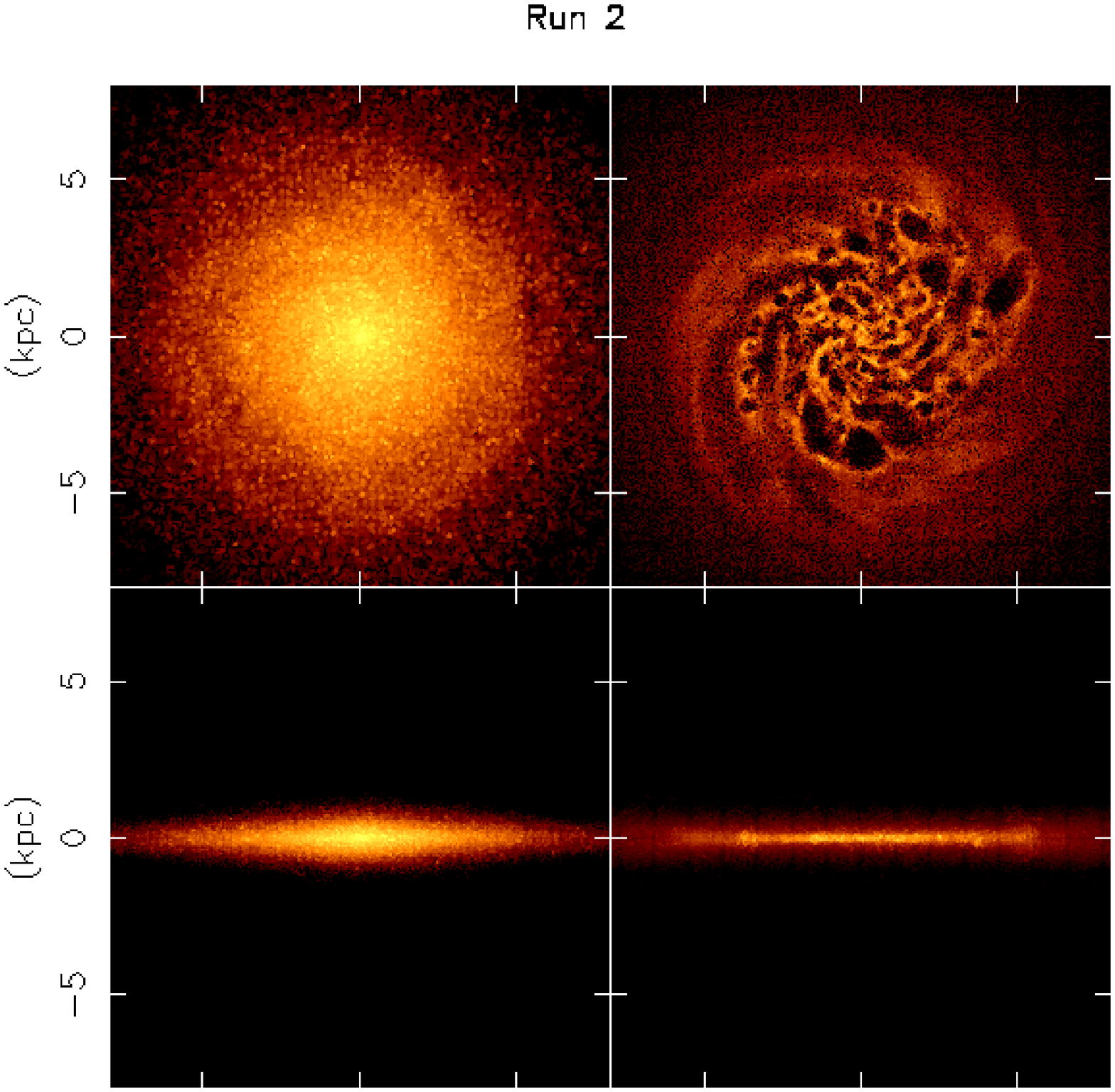}} \\
\subfloat{\label{fig:Run3}\includegraphics[scale=0.34]{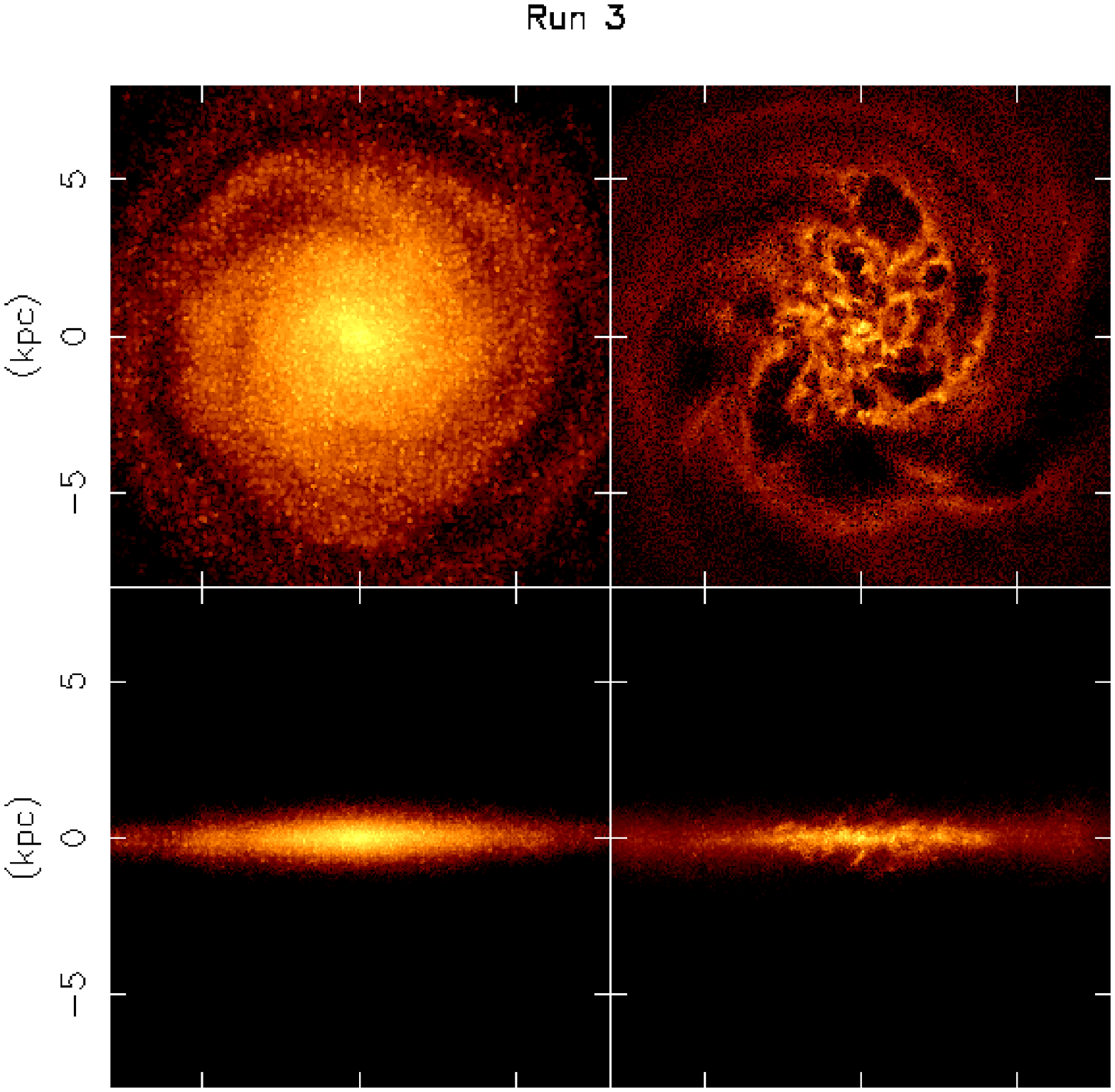}} \hspace{5.0mm}  
\subfloat{\label{fig:Run4}\includegraphics[scale=0.34]{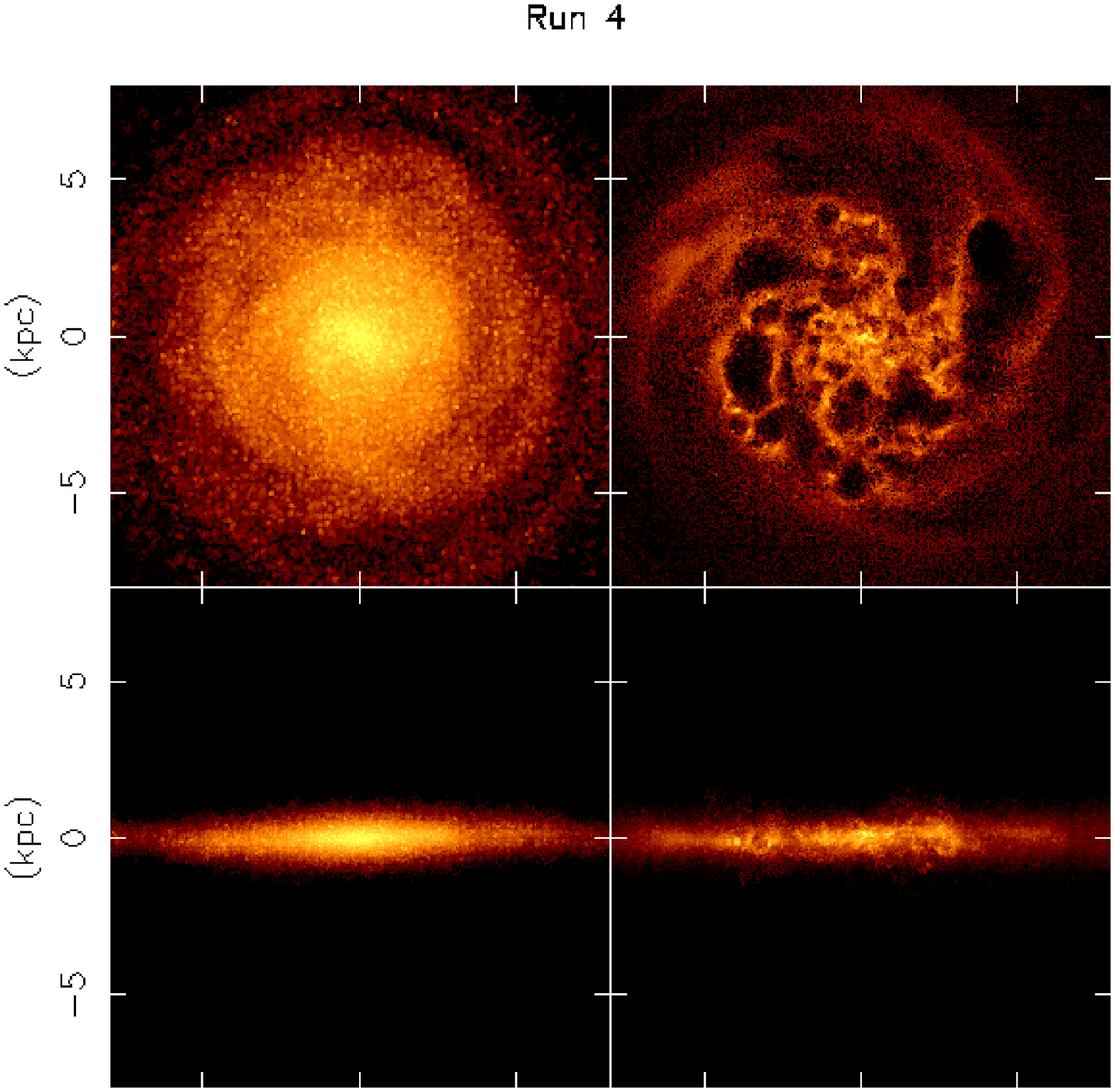}} \\
\subfloat{\label{fig:Run5}\includegraphics[scale=0.34]{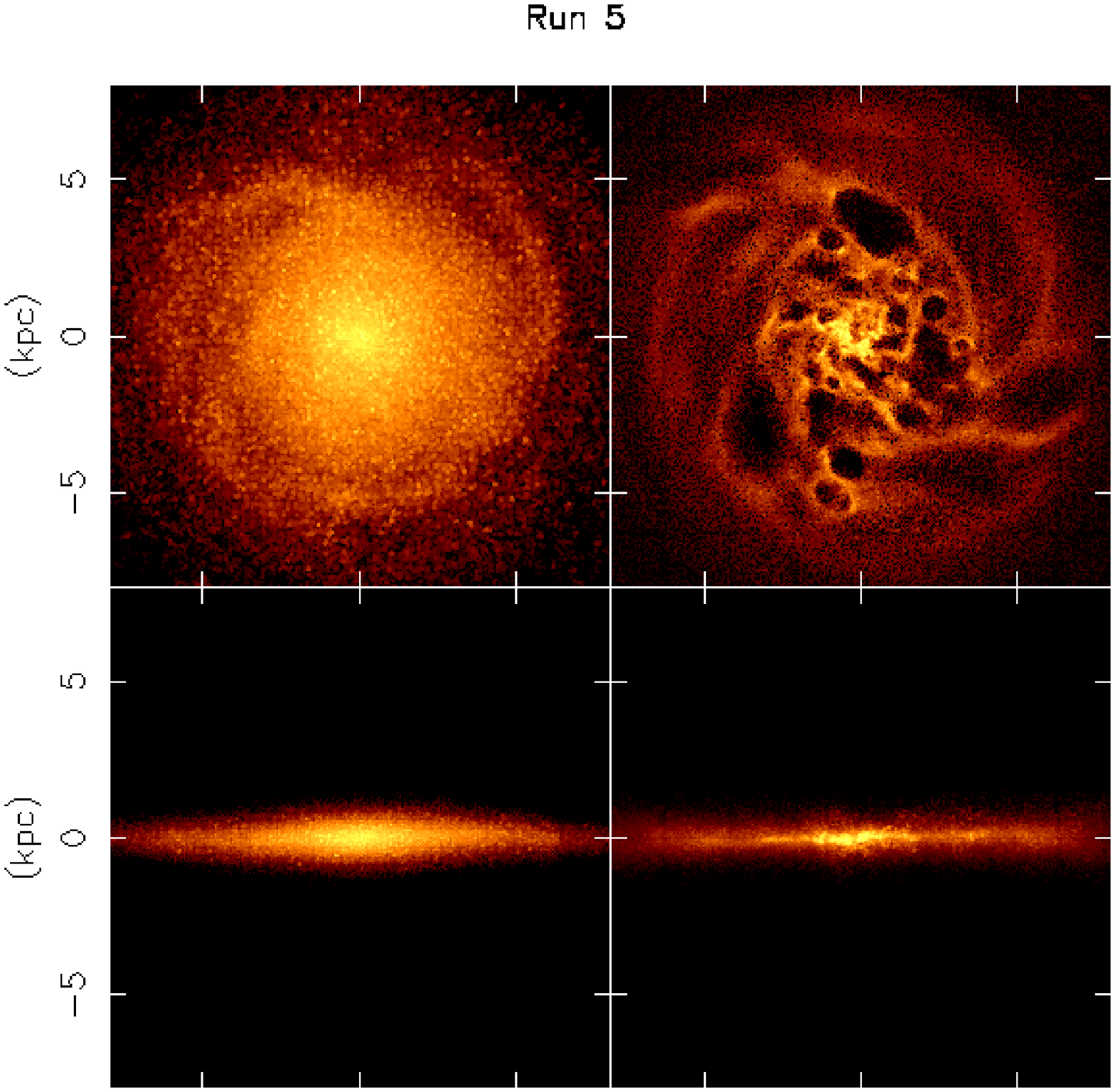}} \hspace{5.0mm}  
\subfloat{\label{fig:Run6}\includegraphics[scale=0.34]{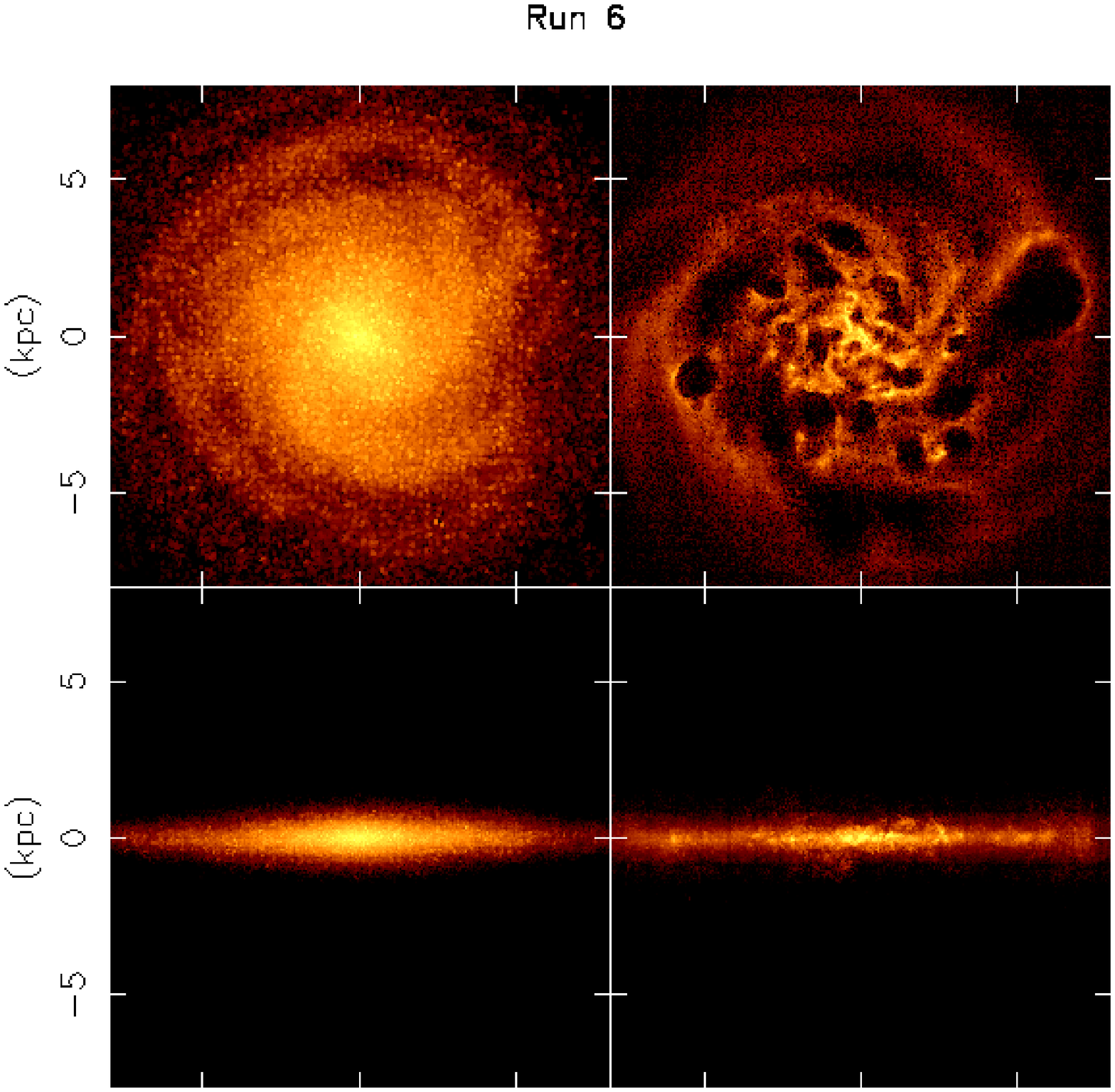}} \\
 \caption{Face on (top row) and edge on (bottom row) view of the stellar (left column) and gas (right column) distribution for our six simulation runs at t = 1 Gyr when we carry out our analyses. The large holes in the gas distribution are caused by supernovae. The holes are larger in Runs 3$-$6 since $E_{\rm SN}$ is ten times higher in these runs compared to Runs 1 and 2. Consequently the gas discs of Runs 1 and 2 are razor thin compared to Runs 3$-$6 which  have a larger vertical gas velocity dispersion.}
 \label{fig:snapshots}
\end{figure*}  
   
\subsection{Model galaxy setup}
\label{model-setup}

 We are interested in investigating the evolution of a late-type disc galaxy of the order of size of M33. We therefore set up an isolated disc galaxy which consists of gas and stellar discs with no bulge component in a static dark matter halo potential. 
We use the standard Navarro-Frenk-White (NFW) dark matter halo density profile \citep{nfw97}, assuming  a standard cold dark matter ($\Lambda$CDM) cosmological model with cosmological parameters of $\Omega_0=0.266$, $\Omega_{\rm b}=0.044$ and $H_0=71{\rm kms^{-1}Mpc^{-1}}$:

\begin{equation}
\rho _{dm}=\frac{3H_{0}^{2}}{8\pi G}(1+z_{0})^{3}\frac{\Omega _{0}}{\Omega (z)}\frac{\rho _{c}}{cx(1+cx)^{2}},
\end{equation}
where
\begin{equation}
c=\frac{r_{200}}{r_{s}}, \;\; x=\frac{r}{r_{200}},
\end{equation}
and
\begin{equation}
r_{200}=1.63\times 10^{-2}(\frac{M_{200}}{h^{-1}M_{\odot }})^{\frac{1}{3}}[\frac{\Omega _{0}}{\Omega (z)}]^{-\frac{1}{3}}(1+z_{0})^{-1}h^{-1} \textup{kpc},
\end{equation}

\noindent where $\rho _{c}$ is the characteristic density of the profile, $r$ is the distance from the centre of the halo and $r_{s}$ is the scale radius. The halo mass is set to be $M_{200}=5.0\times 10^{11}M_{\odot }$ and the concentration parameter is set at $c=10$ \citep{rk08}.

The stellar disc is assumed to follow an exponential surface density profile:

\begin{equation}
\rho _{d}=\frac{M_{d}}{4\pi z_{d}R_{d}}\textup{sech}^{2}(\frac{z}{z_{d}})\exp -({\frac{R}{R_{d}}}),
\end{equation}
\noindent where $M _{d}$ is the disc mass, $R_{d}$ is the scale length and $z_{d}$ is the scale height. Following suggestions from observations of M33 and carrying out several test runs at lower resolution, we decided to use $M _{d}=4.0\times10^{9}$ M$_{\odot}$, $R_{d}=1.4$ kpc and $z_{d}=0.3$ kpc. 
A summary of the numerical values used for our model parameters in the initial model setup is given in Table 1. We used $4\times10^{5}$ stellar particles in our simulations giving each star particle a mass of $1\times10^{4}$ M$_{\odot}$.

\begin{figure}
\centering
\includegraphics[width=\hsize]{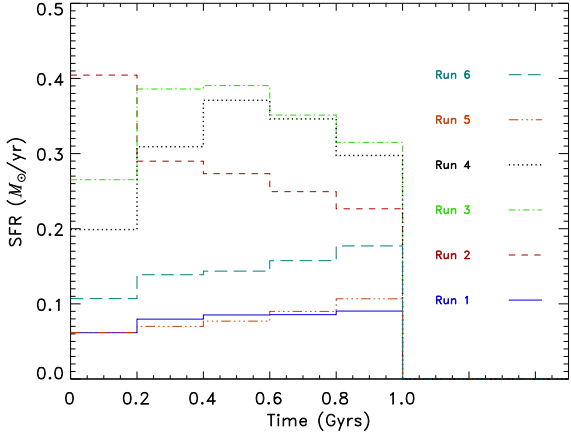}
\caption{
Star formation rate (SFR) history of the simulations. The solid, dashed, dot-dashed, dotted, triple dot-dashed and long-dashed lines represent Runs 1, 2, 3, 4, 5 and 6 respectively.} 
\label{totsfr-fig}
\end{figure}

The gaseous disc is set up following the method described in \citet{sdh05b}.
The radial surface density profile is assumed to follow an exponential law like the stellar disc with a scale length, $R_{d,gas}$ almost three times larger than the stellar disc. The initial vertical distribution of the gas is iteratively calculated to reach hydrostatic equilibrium assuming the equation of state calculated from our assumed cooling and heating function. The total gas mass is $2.0\times 10^{9}$ M$_{\odot }$ and is equal to half the total stellar mass. Since we use $2\times10^{5}$ gas particles in each run, each SPH particle also has a mass of $1\times10^{4}$ M$_{\odot}$. Note that this means both the star and gas particles have the same mass resolution in our simulations. This is not by chance but required from our new modelling of star formation, feedback and metal diffusion shown in Section 2.1. We apply the spline softening and variable softening length suggested by \citet{pm07} for SPH particles. We set the minimum softening length at 198 pc, which corresponds to the required softening for gas with solar metallicity and density of  $n_{\rm H}=1$ cm$^{-3}$, i.e. the assumed density threshold for star formation. For the star particles, we applied a fixed spline softening of 198 pc softening length. We summarise our simulation properties in Table 1. Column 1 represents the NFW halo concentration parameter. The second column represents the virial mass; the third column is the virial radius; Columns 4 and 5 represent the mass of the stellar and gas disc respectively. Columns 6 and 7 represent the scale lengths of the stellar and gas discs respectively and column 8 is the scale height for the stellar disc.

\section{Results}
\label{res-sec}

\begin{figure}
\centering
\includegraphics[angle=0,width=\hsize]{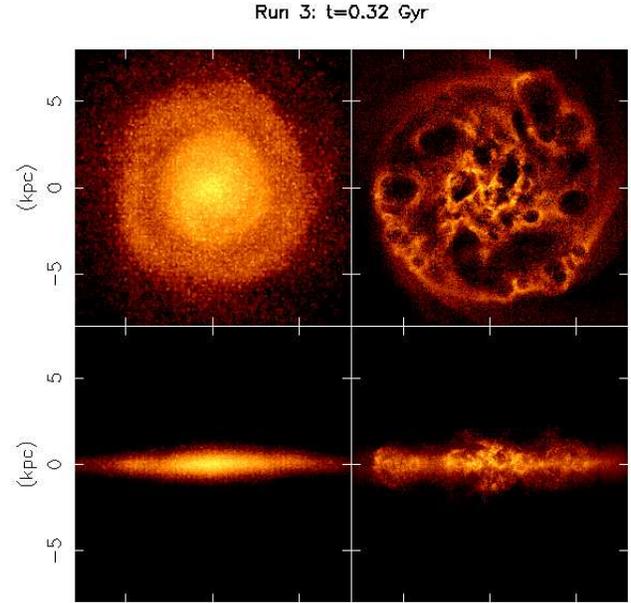}
\caption{
Face-on (upper) and edge-on (lower) snapshot of the galaxy at t = 0.32 Gyr for stars (left column) and gas (right column) colour coded by density for Run 3. The brightest regions are the densest. Large holes or bubbles are visible in the gas distribution due to powerful SNe explosions. The large scale vertical turbulence this causes is visible in the edge-on gas distribution.
}
\label{luminmap-fig}
\end{figure}

Using the galaxy simulation parameters shown in Table 1, we carried out six primary simulation runs, named Runs 1-6 where we consecutively changed one extra model parameter between Runs 1-4. Run 5 is for comparison with Runs 1 and 4 and Run 6 is our fine tuned `best model'. Our varying model parameters include $C_{\rm *}$, $E_{\rm SN}$ and $E_{\rm SW}$. Run 1 has low values for all three of our model parameters. We carried out a simulation run (Run 2) with $C_{\rm *}$=0.2, ten times higher star formation efficiency. We shall refer to this model as the  `high-$C_{\rm*}$ model'. Another run (Run 3) was done with $C_{\rm *}$=0.2 and $E_{\rm SN}$=$1.0\times10^{51}$ ergs. We call this run `high-$E_{\rm SN}$ model'. In our fourth model run, we also increase $E_{\rm SW}$ by a factor of ten. We call this model our `highest-feedback model'. Run 5 has the same high feedback parameters as in Run 4, but with a low $C_{\rm *}$ value as in Run 1. Run 6 is identical to Run 5, except that we increase the value of $C_{\rm *}$ from 0.02 to 0.05. We chose this value based on the results of the previous runs, believing this to be our best model.
Table 2 summarises the simulation runs and model parameters.

Fig.~\ref{fig:snapshots} shows face-on and edge-on snapshots of the stars and gas after 1 Gyr of evolution for our six runs. Large holes or `bubbles' of varying sizes are clearly visible in the face-on gas distributions. These bubbles are produced mainly by  SNe II explosions. The bubbles are much larger in Runs 3$-$6 since the energy released per supernova explosion is ten times greater in these models. The large bubbles in these models may survive for longer than one galactic  rotation. The models with high feedback and star formation efficiency parameters (Runs 3 and 4) produce the most turbulent vertical gas distribution as can be seen in the edge-on images. Conversely, the vertical gas distribution of the discs of Runs 1 and 2 are razor thin in comparison. Clearly then SNe play a critical role in shaping the ISM properties. We shall look into this more quantitatively later. 
The vertical gas distribution of Runs 5 and 6 are slightly less turbulent (compared to Runs 3 and 4) since these runs employ a significantly lower $C_{\rm *}$ value. 

\begin{figure}
\centering
\includegraphics[width=\hsize]{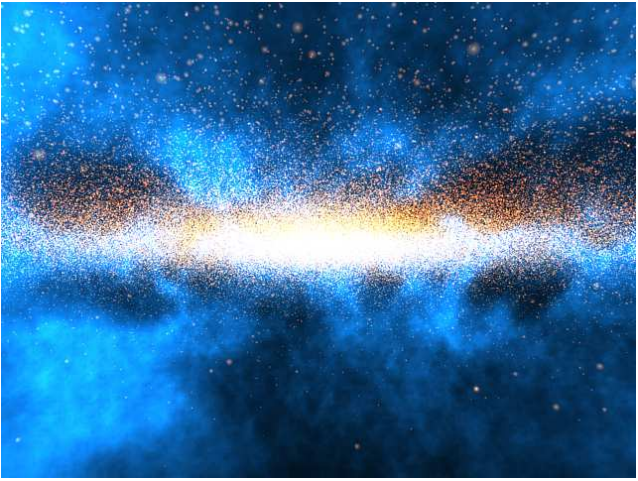}
\caption{Three-dimensional visualisation of the early chaotic scenes in one of our high feedback  simulations (Run 4). This snapshot was taken with the observer embedded in the plane of the disc looking toward the centre of the galaxy. The yellow spheres represent the stars and the blue (diffuse) material is the gas. The colour coding is by density, so whiter or bluer regions represent more dense conglomerations of stars and gas respectively. The clumpy and highly inhomogeneous nature of the gas is clearly visible, as are the holes punched out by SNe. The larger stars are closer to the observer.} 
\label{3d-fig}
\end{figure}

\begin{table}
 \centering
 \begin{minipage}{140mm}
  \caption{Model parameters}
  \renewcommand{\footnoterule}{}
  \begin{tabular}{@{}lllllllllll@{}}
  \hline
   Model & $C_{\rm *}$ & $E_{\rm SN}$ & $E_{\rm SW}$ \\
   & & $(\textup{erg})$ & $(\textup{erg s}^{-1})$ \\
   \hline
Run 1 & 0.02 & $1.0\times10^{50}$ & $1.0\times10^{36}$ \\
Run 2 & 0.2 & $1.0\times10^{50}$ & $1.0\times10^{36}$ \\
Run 3 & 0.2 & $1.0\times10^{51}$ & $1.0\times10^{36}$ \\
Run 4 & 0.2 & $1.0\times10^{51}$ & $1.0\times10^{37}$ \\
Run 5 & 0.02 & $1.0\times10^{51}$ & $1.0\times10^{37}$ \\
Run 6 & 0.05 & $1.0\times10^{51}$ & $1.0\times10^{37}$ \\
 \hline
\end{tabular}
\end{minipage}
\end{table}

\begin{figure}
\centering
\includegraphics[angle=0,width=\hsize]{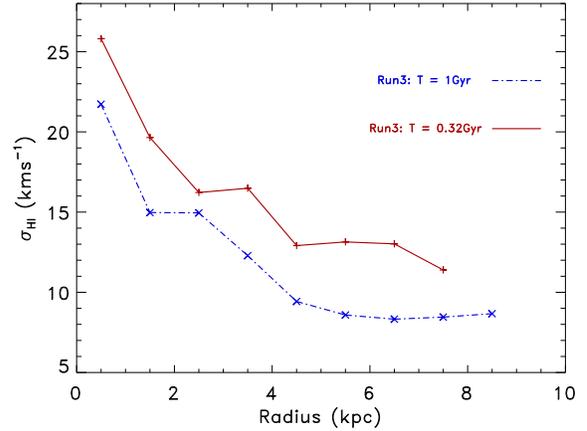}
\caption{
Vertical gas velocity dispersion against radius seen at two different times in the same simulation (Run 3). The solid red line shows the result at an early and chaotic time whereas the blue dot-dashed line shows the velocity dispersion at much later times. During the large bubble phase, the velocity dispersion is unrealistically high.
}
\label{vd1-fig} 
\end{figure}

Fig.~\ref{totsfr-fig} shows the total SFR as a function of time for our different simulation models.
From Fig.~\ref{totsfr-fig} we see that model Runs 1 and 5 have the lowest SFR. This is due to their low $C_{\rm *}$ value. Runs 2, 3 and 4 have a global SFR varying between 0.2 - 0.4 M$_{\odot}\textup{yr}^{-1}$ over 1 Gyr of evolution. There are important differences between these models however. Model Run 2 starts with a initially high SFR at 0.4 M$_{\odot}\textup{yr}^{-1}$ and decreases fairly consistently towards 0.2 M$_{\odot}\textup{yr}^{-1}$ after 1 Gyr. This is because the high star formation efficiency of this model means an initially high SFR, but as time passes and the gas reservoir from which the stars are formed quickly gets used up, the SFR correspondingly  decreases. Runs 3 and 4 initially have a lower SFR compared to Run 2. This is due to large outflows or `bubbles' which drive out the gas at very early times. After this initial violent period the SFR increases and peaks after 0.5 Gyrs at 0.4 M$_{\odot}\textup{yr}^{-1}$ after which it decreases slightly to 0.3 M$_{\odot}\textup{yr}^{-1}$ by 1 Gyr of evolution.
This means the higher energy produced from SNe and stellar winds initially slightly suppress star formation at very early times but that more gas becomes available for later star formation.
The early time suppression of star formation is slightly stronger in Run 4 compared to Run 3 since in this model we implement higher energy release from stellar winds as well as from SNe. Run 5 has identical feedback parameters to Run 4, with the only difference being that $C_{\rm *}$ is low as in Run 1. Run 5 has a similar SFR history to Run 1, showing that the SFR is extremely sensitive to the star formation efficiency ($C_{\rm *}$) as expected. Run 6 has roughly double the SFR of Run 5 as expected from its higher $C_{\rm *}$ value. 
After approximately 0.5 Gyrs the SFR does not evolve significantly in all our simulation runs. Before 0.5 Gyrs the ISM can be extremely turbulent, especially for the high-feedback models, as can be seen in Fig.~\ref{luminmap-fig} which shows the 2-D face-on and edge-on stellar and gas distribution at an early time (0.32 Gyrs) for our high-$E_{\rm SN}$ model (Run 3). At this time, many large gas outflows  are seen to occur. Fig.~\ref{3d-fig} shows a 3-D visualisation of both the stellar and gas distributions of the galaxy at early times for Run 4. The clumpy and chaotic nature of the ISM is clearly apparent at these early times, with large gas outflows and bubbles visible. Fig.~\ref{vd1-fig} shows that during the time when a large gas outflow occurs, the velocity dispersion of the gas component (measured as described below) is correspondingly very high. At a much later time when the disc is more settled (e.g. 1 Gyr), the velocity dispersion is correspondingly lower. Therefore 1 Gyr was chosen as a sufficiently long enough time to wait for the system to settle before carrying out our main analysis and comparison to the observations.

\begin{figure*}
\centering
\includegraphics[width=\hsize]{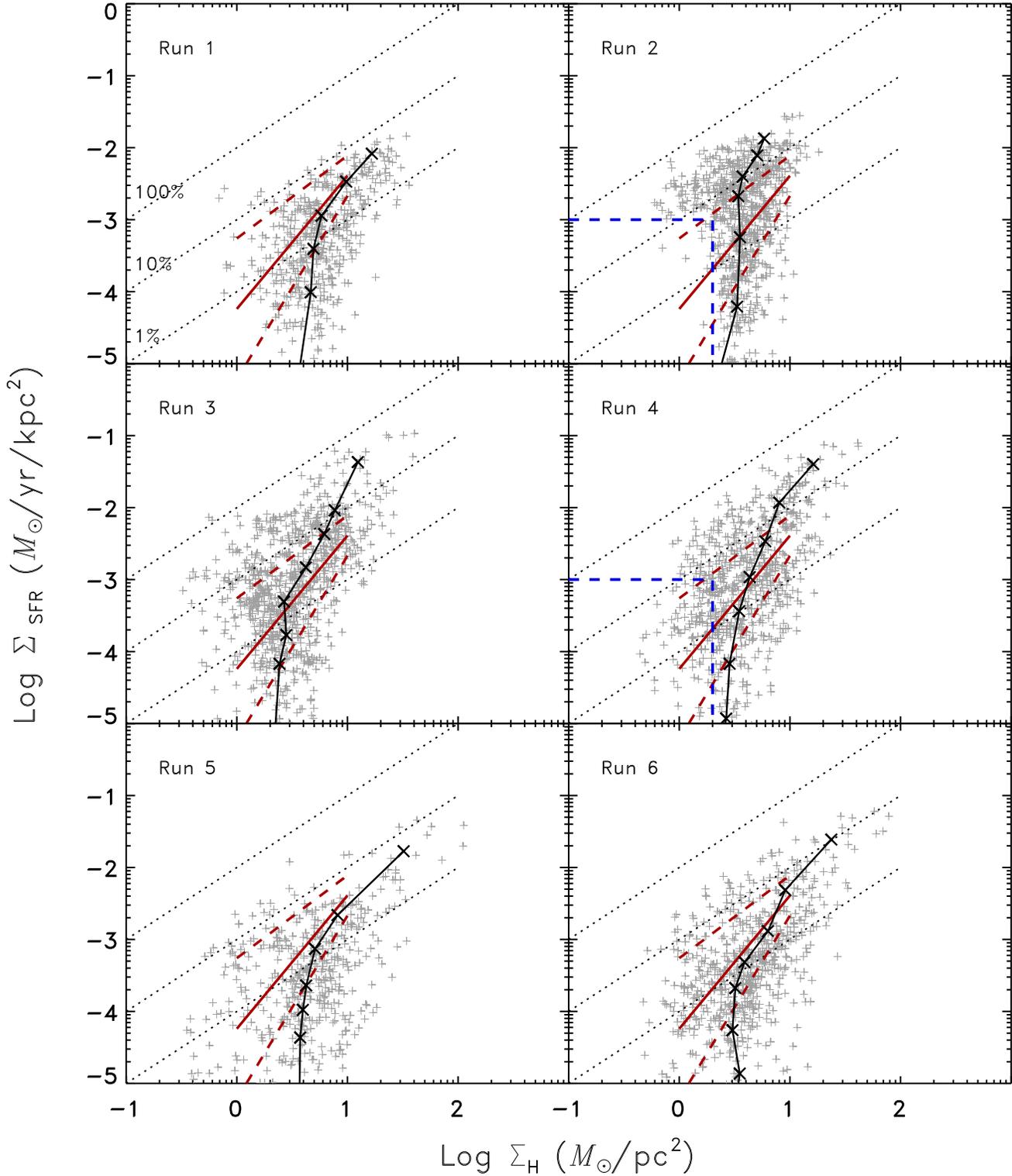}
\caption{
Surface SFR density as a function of surface gas density for our six models at 333 pc resolution (grey plus signs). The surface SFR densities are calculated using the surface densities of young stars less than 0.5 Gyrs old. The diagonal dotted lines represent lines of constant SFE, indicating the level of  
$\Sigma_{\rm SFR}$ needed to consume 1, 10 and 100 per cent of the gas reservoir in $10^{8}$ years, exactly as in fig. 13 of \citet{blw08}. Therefore the lines also correspond to constant gas depletion times of  $10^{10}$, $10^{9}$ and $10^{8}$ yr respectively. The red solid and dashed lines represent the mean and 1 sigma upper and lower bounds respectively for the \citet{blw08} data between log $\Sigma_{\rm H}$ values of 0 and 1. The blue dashed horizontal and vertical lines in Run 2 are drawn to emphasize the region of low $\Sigma_{\rm H}$ and $\Sigma_{\rm SFR}$ where there are barely any pixels present. The same region is marked in Run 4 to show that this model can successfully produce regions with both low $\Sigma_{\rm H}$ and $\Sigma_{\rm SFR}$. The location of these pixels are shown in Fig.~\ref{bubblepixel-fig}. In this figure $\Sigma _{\rm H}=\Sigma _{\rm HI}+\Sigma _{\rm H_2}$. 
We also overplot the results obtained using azimuthally averaged ring annuli of width 1 kpc (black crosses connected by solid lines). 
}
\label{SK-fig}
\end{figure*}

\begin{figure}
\centering
\includegraphics[angle=0,width=\hsize]{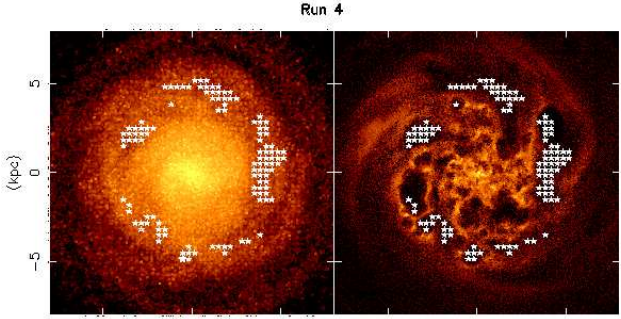}
\caption{
Face on view of the stellar (left) and gas (right) distribution at t = 1 Gyr for model Run 4. The white stars represent the spatial locations of the pixels  selected in Run 4 of Fig.~\ref{SK-fig} which contain both low surface gas and SFR densities (enclosed by the dashed blue lines and the axes). We can clearly see that all of these pixels are located in bubble regions where the gas density is low.
}
\label{bubblepixel-fig} 
\end{figure}

Fig.~\ref{SK-fig} shows the relation between gas and SFR surface densities for our six sample runs where we have plotted the units to be identical to the observed sample of spiral galaxies in fig. 13 of \citet{blw08}. 
\citet{blw08} show the surface SFR density against total hydrogen surface density. Taking advantage of our self-consistent chemodynamical modelling, we analyse the hydrogen fraction for each gas particle and self-consistently show the total hydrogen surface density in Fig.~\ref{SK-fig}, where in our notation: $\Sigma _{\rm H}=\Sigma _{\rm HI}+\Sigma _{\rm H_2}$.
The results were obtained using a pixel-by-pixel methodology using a square-cell resolution of 333 pc x 333 pc and a smoothing length of two pixel box lengths. The surface SFRs are calculated using the surface densities of young stars less than 0.5 Gyrs old. We only include the pixels which have at least 10 young star particles contributing to them. 
In Fig.~\ref{SK-fig}, we also plot diagonal dotted lines of constant star formation efficiency (SFE), exactly as in \citet{blw08} in order to make the comparison easier, indicating the level of $\Sigma_{\rm SFR}$ needed to consume 1, 10 and 100 per cent of the gas reservoir in $10^{8}$ years. We also overplot a best fit line (solid red line) to represent the mean values for the \citet{blw08} data between log $\Sigma_{\rm H}$ values of 0 and 1 \citep[see e.q. 2 of][]{blw08} and show the 1 sigma upper and lower bounds (dashed red lines). We can clearly see that the $\Sigma_{\rm SFR}$ for model Run 1 is too low at all values of $\Sigma_{\rm H}$. This is because this simulation had a very low star formation efficiency. Increasing the SFE parameter (Run 2) does increase $\Sigma_{\rm SFR}$ but at the expense of accurately following the slope of the observations. Our high-$C_{\rm*}$ model (Run 2) therefore still fails to accurately follow the observed trends.
The high-$E_{\rm SN}$ model (Run 3) and our highest feedback model (Run 4) better follow the observed slope and values for the SK law. Interestingly, the increased energy released from SNe explosions and stellar winds in these models help to increase the star formation rate at very low gas densities. 
The major difference between Run 2 and Run 4 comes from the lack of pixels with low values for both $\Sigma_{\rm H}$ and $\Sigma_{\rm SFR}$ in Run 2, unlike the higher feedback models such as Run 4.
By selecting the pixels that fall into the bottom-left region (enclosed by the dashed blue lines) for Run 4 in Fig.~\ref{SK-fig}, 
 we investigated where these pixels were located and found that they predominantly lie in large holes blasted out by SNe (Fig.~\ref{bubblepixel-fig}). This suggests that SNe may play a vital role in shaping observed SK-laws, especially at low gas densities.
Run 5 also employs high feedback parameters, but with a low $C_{\rm *}$ value of 0.02 as in Run 1. Correspondingly, we see that $\Sigma_{\rm SFR}$ is too low like in Run 1. In Run 6 we have increased the value of $C_{\rm *}$ to 0.05. This represents our best compromised model. We see that increasing $C_{\rm *}$ results in higher $\Sigma_{\rm SFR}$ values and a better fit to the observations. We note that due to the relatively low mass of our simulated galaxy, which does not have a very high SFR or contain very dense regions, we cannot inspect a wide range of $\Sigma_{\rm H}$ parameter space and are therefore limited to making our comparisons in a narrow range of  $\Sigma_{\rm H}$. 

In Fig.~\ref{SK-fig} we also overplot the results obtained using a radial ring analysis method (black crosses connected by solid lines) whereby we have calculated the values of $\Sigma_{\rm H}$ and $\Sigma_{\rm SFR}$ in azimuthally averaged ring annuli of width 1 kpc. We see that both the radial ring and pixel-by-pixel methods produce broadly consistent results, with the possible exception being that the radial ring method does not probe well the low $\Sigma_{\rm H}$ regime. Some of the data points in this regime represent ring annuli regions lying just outside the main star forming disc of our galaxy.

\begin{figure*}
\centering
\includegraphics[angle=0,width=\hsize]{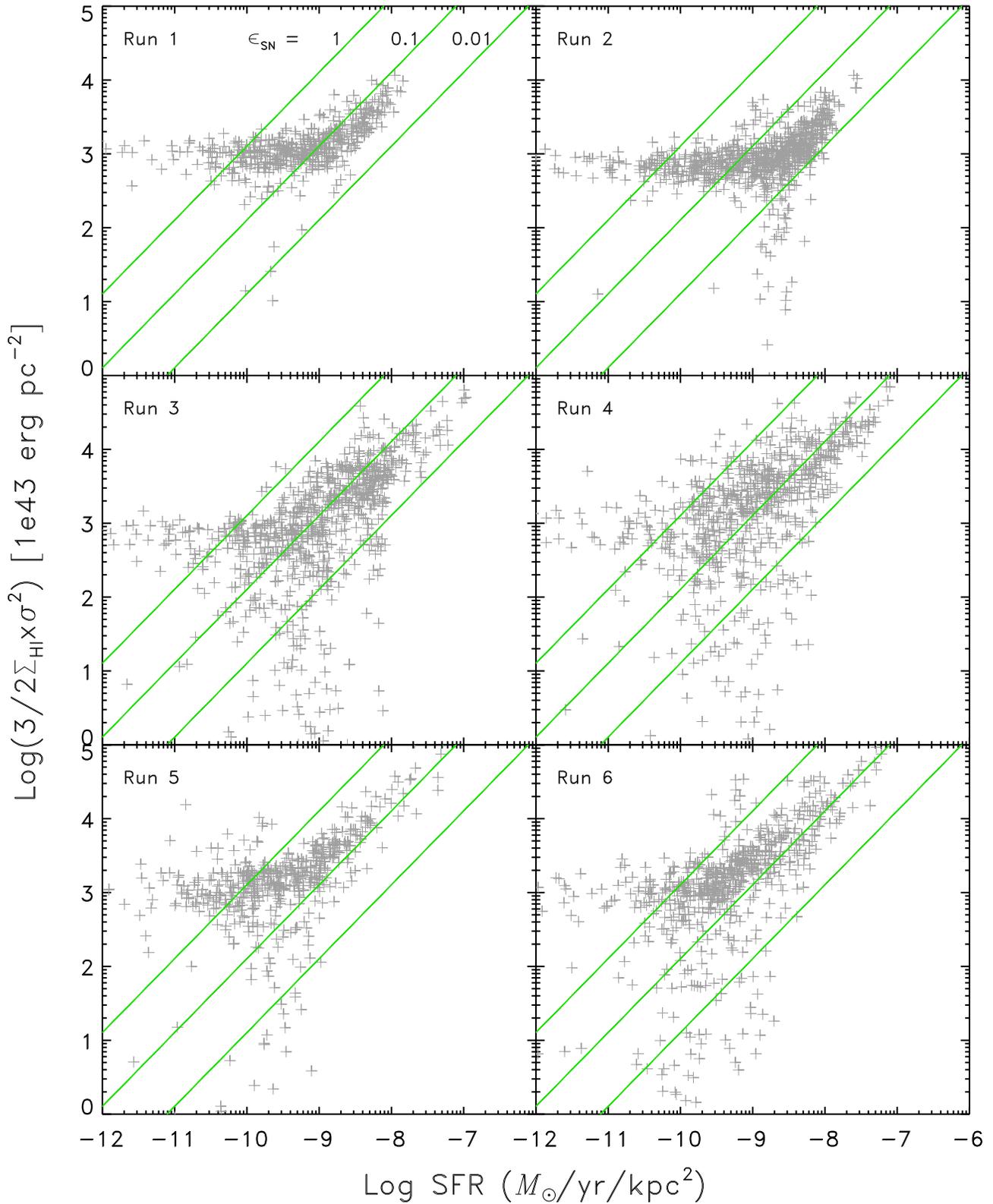}
\caption{
Pixel-by-pixel scatter plot of kinetic energy vs. surface star formation rate for our six model galaxy runs (grey plus signs). The solid lines of unity slope represent the supernova energy input for different supernova efficiency values, $\epsilon _{\rm SN}=1,0.1,0.01$ (top to bottom).
}
\label{KE-fig}
\end{figure*}

In Fig.~\ref{KE-fig} we compare the kinetic energy density of the HI gas, $E_{k}=(3/2)\Sigma _{\rm HI}\sigma _{\rm HI}^{2}$, against the SFR surface density as in \citet{tam09}. 
Here $\sigma _{\rm HI}=\sqrt{\sigma _{\rm th}^{2}+\sigma _{\rm t}^{2}}$ where $\sigma _{\rm th}$ and $\sigma _{\rm t}$ are the velocity dispersions due to thermal and turbulent motions respectively.
In addition, to mimic HI observations, we only take into account the gas particles whose density is in the region $0.1<\rho_{\rm HI}<50$ cm$^{-3}$ and have temperature less than $10^{4}$ K.
We also use the same units for the axes and show with solid lines the supernova energy input for  different SNe efficiency values, $\epsilon _{\rm SN}=1,\: 0.1$ and 0.01 exactly as in  \citet{tam09}. 
Their observations show a linear correlation between $E_{k}$ and $\Sigma_{\rm SFR}$, with typical $\epsilon _{\rm SN}$ values between 0.1$-$1. 
 At higher $\Sigma_{\rm SFR}$ regimes, the $E_{k}$ slightly falls off. Comparing against our simulated galaxies with their Fig. 4, we see that Runs 3$-$6 best follow the observations. Runs 1 and 2 fail to capture the observed trends, especially, $E_{k}$ is lower at $-9.5<\textup{Log}\: \Sigma_{\rm SFR}<-8.5$ compared to Fig. 4 of \citet{tam09}.

\begin{figure*}
\centering
\includegraphics[angle=0,width=\hsize]{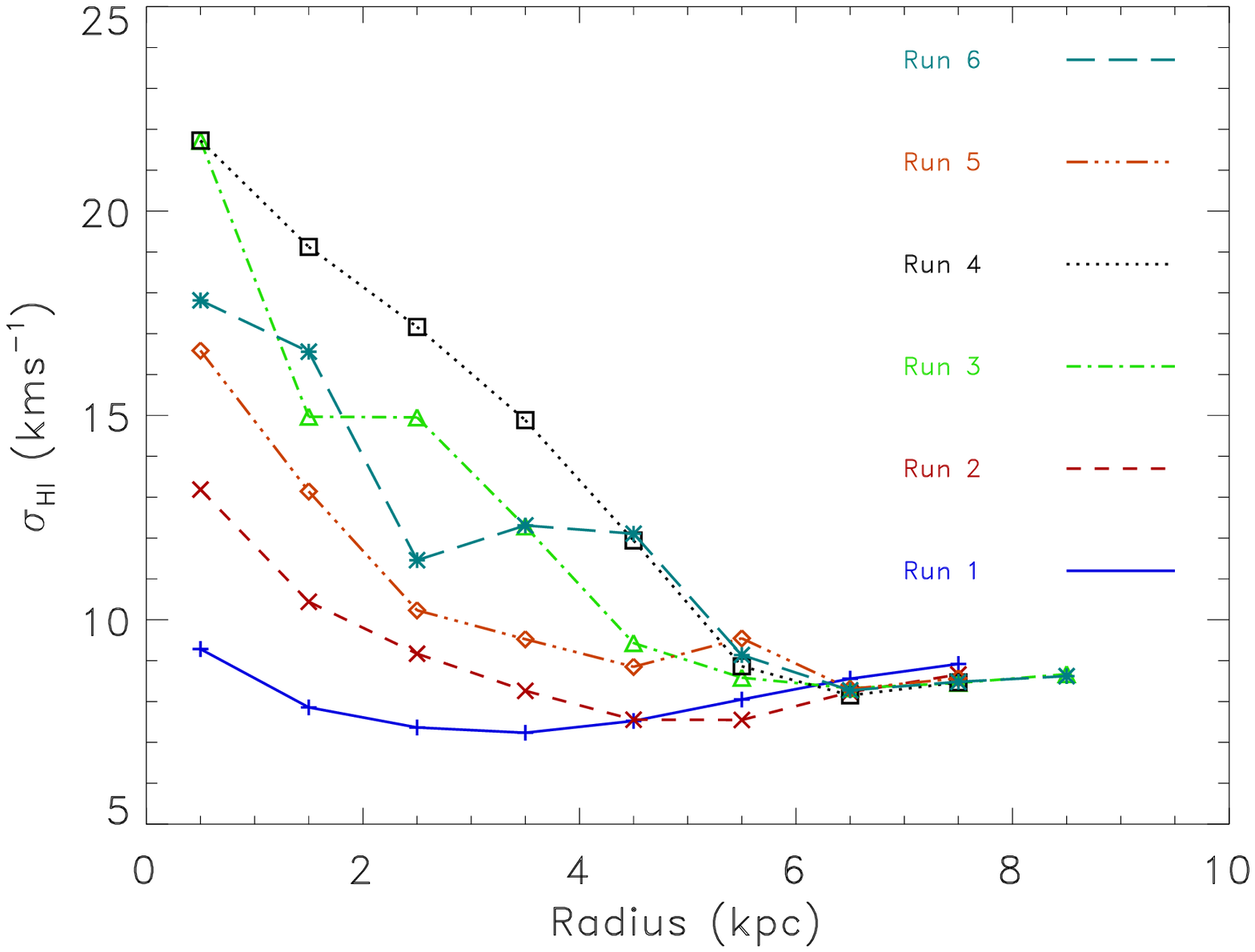}
\caption{
Gas velocity dispersion versus radius for our six models. The `higher-feedback' with high-$C_{\rm *}$ models (Runs 3 and 4) have larger velocity dispersion. Our results show that the velocity dispersion is primarily sensitive to our SNe energy ($E_{\rm SN}$) parameter.
}
\label{VD2-fig}
\end{figure*} 

We also analysed the azimuthally averaged radial profile for the HI velocity dispersion, by calculating the average value of $\sigma _{\rm HI}$ (following the same method as above) within ring annuli of radial width 1 kpc.
Observations of spiral galaxies show a radial decline for the HI velocity dispersion with a value of $\sim$ 20 km s$^{-1}$ in the inner region, declining to $\sim$ 10 km s$^{-1}$ by $r_{25}$ \citep{tam09}. Fig.~\ref{VD2-fig} shows that our higher feedback models (Runs 3 and 4) closely follow the aforementioned observed trends. All of our  models had similar $\sigma _{\rm HI}$ values of around 10 km s$^{-1}$ by $r_{25}$ similar to what was found in \citet{tam09}. With the exception of Run 1, all models showed a radial decline of $\sigma _{\rm HI}$. The decline was small for our high-$C_{\rm*}$ model (Run 2)  and more noticeable for our higher feedback models (Runs 3 and 4). This is understandable since the higher feedback models produce larger gas outflows and bubbles which increases $\sigma _{\rm HI}$. Interestingly, our model with high $E_{\rm SN}$ and $E_{\rm SW}$ (Run 4) did not have a noticeable difference with our high-$E_{\rm SN}$ model (Run 3), proving that it is the supernova energy which mainly drives the turbulence in the ISM in our models. Further evidence of this lies in the comparison of the velocity dispersion of Runs 1 and 5, with Run 5 having a noticeably larger $\sigma _{\rm HI}$. Despite this, $\sigma _{\rm HI}$ for Run 5 is still too low compared with observations of \citet{tam09}. Therefore in Run 6, we slightly  increased the value of $C_{\rm *}$ and found that $\sigma _{\rm HI}$ also increased to a level which is consistent with the observations.



\section{Summary and Conclusions} 
\label{conc}

We aim to build the most accurate and self-consistent numerical model for late-type spiral galaxies to date. To this end we have been developing our numerical simulation code {\tt GCD+}.
We include new star formation and feedback recipes which allow our simulations to follow the effects of powerful SNe explosions on galaxy evolution.

In this study we set up an isolated M33-sized disc galaxy consisting of gas and stellar discs with no bulge component, in a static dark matter halo gravitational potential. 
We carried out six main simulation runs in which we varied three of our model parameters; the star formation efficiency $(C_{\rm *})$, the thermal energy released per SNe explosion ($E_{\rm SN}$) and the thermal energy released per unit time from stellar winds ($E_{\rm SW}$). We studied the effect of varying these parameters on the Schmidt-Kennicutt Law and vertical gas velocity dispersion versus radius relation comparing to observations from \citet{blw08} and \citet{tam09} respectively. We find that our models with higher energy feedback (Runs 3$-$6)  more closely resemble the observations.
Powerful supernova explosions cause the formation of large holes or bubbles in the gas distribution in the galaxy and stir the ISM turbulence and are thus essential for producing the observed gas velocity dispersions. The gas velocity dispersion is also affected (albeit less strongly) by the star formation efficiency. Our model with high energy feedback from SNe and stellar winds and intermediate star formation efficiency (Run 6) most closely resembled the observations. This run represented our `best compromised model' since we had fine-tuned the star formation efficiency and feedback parameters. 
In our pixel-by-pixel analysis of the star formation law, we found that our model with high star formation efficiency and low energy feedback (Run 2) produced a KS relation with an unrealistically steep slope. Models with low energy feedback had a distinct lack of pixels with both low $\Sigma_{\rm H}$ and $\Sigma_{\rm SFR}$, resulting in a failure to accurately reproduce the observed star formation law. By increasing the energy feedback parameters, $E_{\rm SN}$ and $E_{\rm SW}$ (as in Run 4) we managed to rectify this problem and
found that the pixels which had low values for both $\Sigma_{\rm H}$ and $\Sigma_{\rm SFR}$ were predominantly located inside bubble regions, which were more numerous and larger in size for the higher feedback runs. Our results therefore suggest that SNe can play a pivotal role in shaping the star formation law. We used both a pixel-by-pixel and radial ring method to analyse the star formation law in our galaxies and found that overall, both methods gave broadly consistent results.

In this paper we employed static potentials for the dark matter halo and stellar disc. The main benefits of not using a live halo include the prevention of numerical effects such as artificial disc heating due to the scattering of baryonic particles by more massive dark matter particles which may be an issue in lower resolution cosmological simulations. Our numerical simulations thus do not suffer from these global disc instabilities. However, we underestimate the perturbation from accreting satellites \citep[e.g.][]{pbt11} and globular clusters \citep[e.g.][]{vcf09}.

In this paper we demonstrate that reproducing both the SK relation and the observed gas velocity dispersions puts stronger constraints on the sub-grid modelling for galaxy evolution simulations.
Encouraged by the success of this work,  
we are currently in the process of running higher resolution simulations. This should result in us being able to probe deeper into the origins of the $\Sigma_{\rm GAS}$$-$$\Sigma_{\rm SFR}$ relation. It should also mean that our models will be less sensitive to the employed value of the star formation efficiency, given we could resolve molecular cloud formation processes \citep{sdk08}. After calibrating our code at this higher resolution, we then aim to run new extremely high resolution cosmological simulations. The fully self-consistent nature of the new higher resolution cosmological simulation will be a powerful tool  to unravel the mysteries behind  the formation and evolution of late type spiral galaxies like M33.

\section*{Acknowledgments}

We acknowledge the support of the UK's Science \& Technology
Facilities Council (STFC Grant ST/H00260X/1, ST/F002432/1). 
We acknowledge CfCA/NAOJ and UCL's Legion supercomputer where the numerical computations for this paper were performed.

\bibliographystyle{mn}
\bibliography{dkref}

\label{lastpage}

\end{document}